# Considerations for a Multi-beam Multi-purpose Survey with FAST

Di Li[1], Pei Wang[1], Lei Qian[1], Marko Krco[1], Alex Dunning[2], Peng Jiang[1], Youling Yue[1], Chenjin Jin[1], Yan Zhu[1], Zhichen Pan[1], Rendong Nan[1]

National Astronomical Observatories
Chinese Academy of Sciences

1. National Astronomical Observatories, Chinese Academy of Sciences, Beijing, China; Email: dili@nao.cas.cn
2. CSIRO Astronomy and Space Science, P.O. Box 76, Epping, NSW 1710, Australia

## I. INTRODUCTION

Having achieved 'first-light' right before the opening ceremony on September 25, 2016, the Five-hundred-meter Aperture Spherical radio Telescope (FAST) is being busily commissioned. Its innovative design requires ~1000 points to be measured and driven instead of just the two axes of motion, e.g. Azimuth and Elevation for most of the conventional antennae, to realize pointing and tracking. We have devised a survey plan to utilized the full sensitivity of FAST, while minimizing the complexities in operation the system. The 19-beam L band focal plan array will be rotated to specific angles and taking continuous data streams while the surface shape and the focal cabin stay fixed. Such a survey will cover the northern sky in about 220 full days. Our aim is to obtain data for pulsar search, HI (neutral hydrogen) galaxies, HI imaging, and radio transients, simultaneously, through multiple backends. These data sets could be a significant contribution to all related fields in radio astronomy and remain relevant for decades.

## II. THE FAST PROJECT

The Five-hundred-meter Aperture Spherical radio Telescope (FAST: [1], also see Figure 1) was first proposed as part of a Square Kilometer Array (SKA) concept in early 1990s. FAST represented the so called large antenna, small number (LASN) concepts and was one of the final four proposals considered by the international SKA consortium. Although the winning and current SKA design clearly favored the small antenna large number (SALN) concepts, some of the expected "Moore's law" type saving in producing antennae on industrial scale and dropping costs in computation have not fully materialized. As the largest single-dish in the world, FAST is extremely cost-effective in delivering 10% SKA collecting

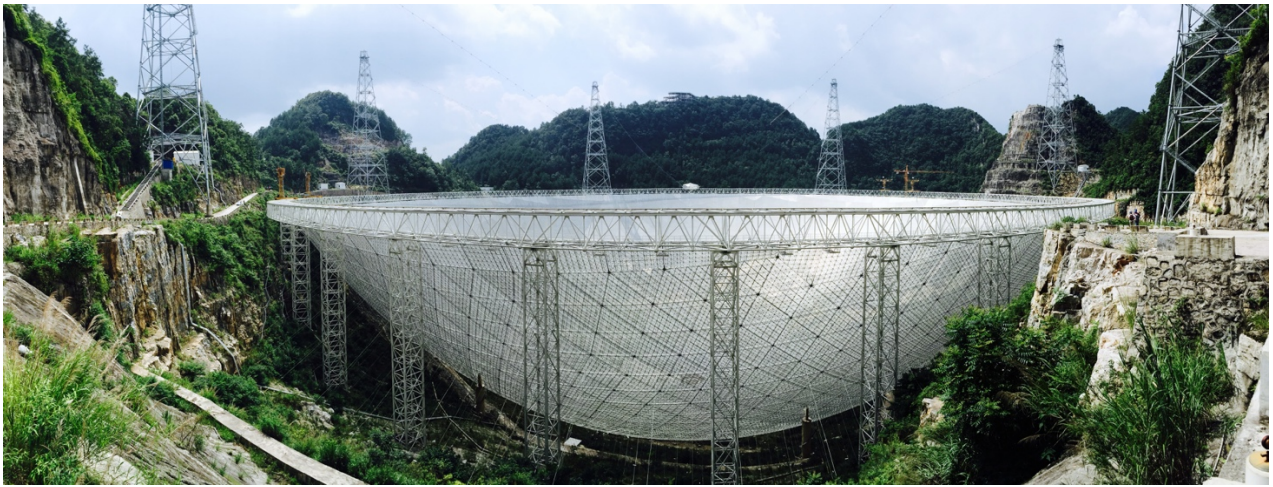

**Fig. 1** *Panoramic view of FAST (obtained in June of 2017).*

area with a modest ¥1.15 billion (RMB) budget.

FAST covers 70 MHz to 3 GHz continuously. At the moment, an ultra-wide receiver (270 MHz to 1.62 GHz) is installed. The 19-beam array receiver for L-band was built by CSIRO and was formally received by NAOC in Marsfield, Sydney in May of 2017. Once the 19-beam arrives at the site, it is expected to occupy the majority of the time for receiver commissioning. The raw sensitivity in L-band should reach 2000 m$^2$ K$^{-1}$. The maximum source-changing was specified as 10 minutes for a maximum 80° slewing. There is a minimum slewing time of about 90 seconds, due to both the settling time of hydraulic actuators and the large number of measurements required. Adopting Table 1 from from ref. [2] with some updates, we list the major specifications of FAST in Table 1.

TABLE I.   MAIN TECHNICAL SPECIFICATIONS OF FAST

Spherical reflector: Radius = 300 m, Aperture = 500 m
Illuminated aperture: $D_{ill}$ = 300 m
Focal ratio: f/D = 0.4611

Sky coverage: zenith angle 26.4° (full gain)  26.4°-40° (with a maximum gain loss of 18%)
Frequency: 70 MHz – 3 GHz
Sensitivity (L-Band): A/T ∽ 2000 m$^2$/K
Resolution (L-Band): 2.9'
Multi-beam (L-Band): beam number = 19
Slewing time: <10 minutes
Pointing accuracy: 8"

## III. THE COMMISSIONING CHALLENGES

The primary surface of FAST consists of 4500 panels, supported by a cable-mesh system, somewhat similar to that of the Arecibo telescope. The control and movement of FAST is unique. More than 2100 actuators anchored to the ground will be actively pulling the connecting nodes and change the tension distribution of cable-mesh system. The typical range of motion for a typical actuator spans roughly 1 meter, but needs to achieve millimeter accuracy. In ref. [3]-[4], Jiang et al. laid out the general strategy of FAST surface adjustment. Figure 1 shows both the expected surface adjustments and measured results from June of 2017.

While the surface is being reshaped, the 30-ton focal cabin hosting receivers will be driven along trajectories constrained to a curved focal plan. Two tier system has been implemented, with the primary control realized by 6 cables connected through 6 towers to 6 ground motor houses and the fine turning "Steward" platform connecting the upper and lower focal cabin.  Figure 2 shows our recent measurements of the focal cabin position during a drift-scan test. No fine tuning was used, just "holding" with 6 cables for the FAST drift-scan mode. The RMS offset mostly stayed around 0.3 cm, with occasional offset shot up to 2 cm due to wind load.

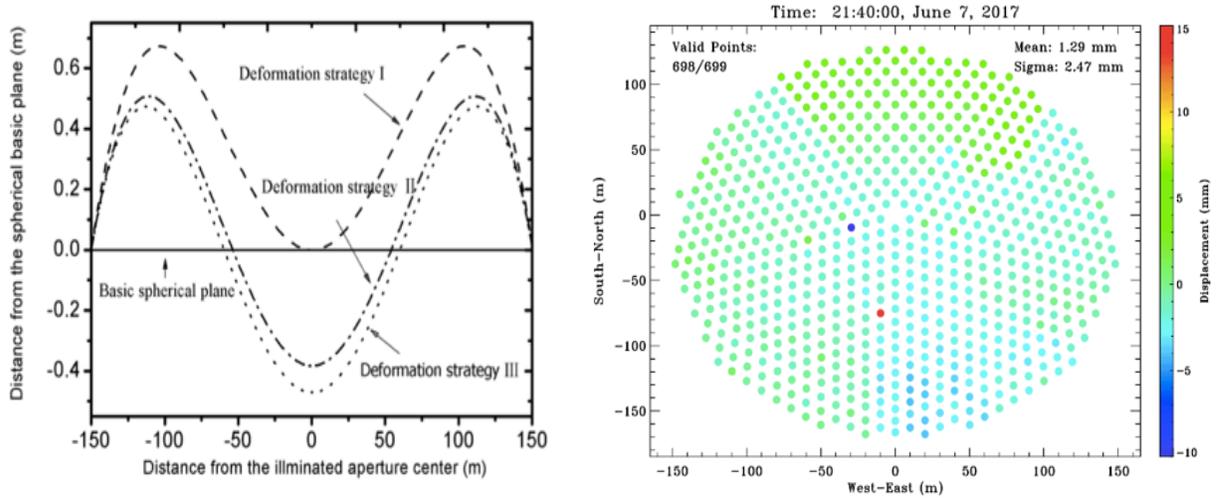

**Fig. 2** *Left panel: the displacement between the default spherical surface of FAST and its intended parabola in operation [3]. Right panel: the measured difference between node positions and the modeled parabola (obtained in June of 2017).*

It is reassuring to see that both the primary surface and the focal cabin can be controlled to accuracy sufficient for L-band observations, without active readjustment and in this relatively early phases of commissioning [5]. The full suite of observing capabilities, however, still present substantial hurdles. Our hydraulic actuators are supposed to hold tons of tension under outdoor conditions, for hundreds of thousands of movements, and maintain mm accuracy. The fault rate of actuator will have to be monitored, modeled, and further improved. The measurement of panel and dome positions are realized by shooting laser beams from 24 measuring towers hosting "total-stations", which are yet to be RFI shielded. During surface adjustment in pointing the telescope, laser beams have to be shot at each individual optical targets one at a time, which also limit the turn over time in knowing the current shape and the ensuing refinement. A real-time simulator has been programed to model the tension distribution and helped optimize the control sequence for any specific observation request [6]. The full suite of planned observation modes, such as tracking, on-the-fly (OTF) mapping, driven scanning, etc., will have to take significantly amount of time to be implemented.

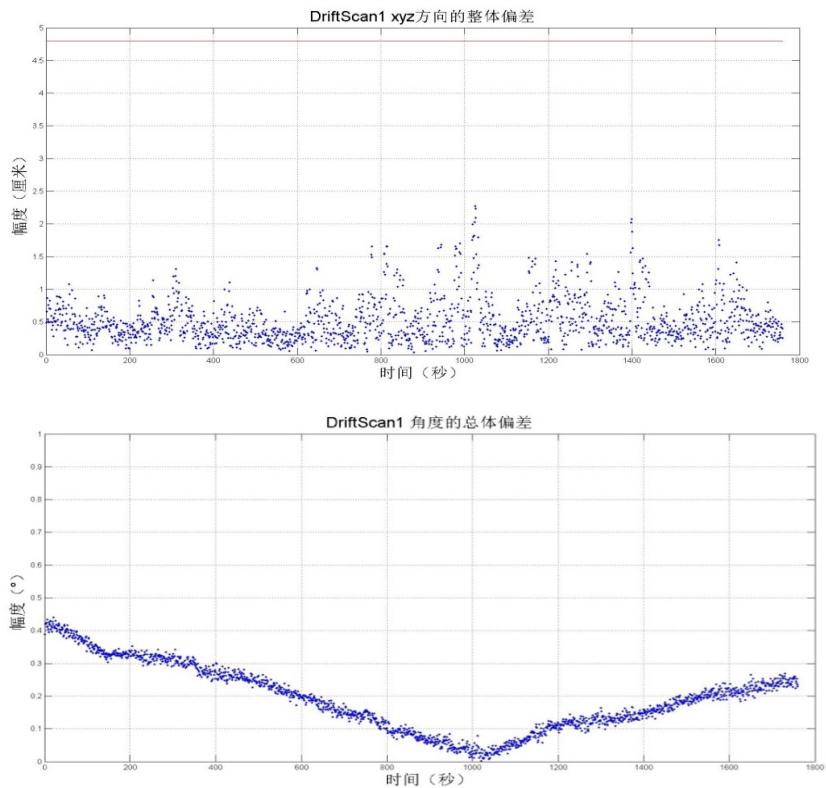

**Fig. 3** *Top panel: the displacement between the measured dome position and and the modeled focal point in centimeters. Bottom panel: the offset angle between the bore sight of the parabola and the dome symmetry axis in degrees. (obtained in June of 2017). The time interval between measurements shown here in seconds and the data span a total of 1500 seconds.*

With the accuracy already achieved for the primary surface and the focal cabin, however, already allows for a drift-scan survey with the full gain of FAST. A drift scan will minimize the complexities in system control and in self-generated RFI. The 19-beam feed-horn array in L-band of FAST is the largest of its kind, following the successful deployment of the Parkes 13-beam array and the 7-beam Arecibo L-band Focal-plane Array (ALFA). With improved design, the FAST 19-beam system achieve better than 20 K receiver temperature over a 400 MHz band covering 1.04 GHz-1.45 GHz. The beam-passing time of a drift scan roughly equals to 15 seconds. With proper weighting, the equivalent integration time per beam will be around 20 seconds. At 25 K system temperature, a FAST drift scan will achieve 0.09 K per 1 km/s RMS noise temperature. For point source, the RMS noise is 0.02 mJy, way lower than the expected confusion limit at about 1 mJy. These expected sensitivities for both spectral line and continuum sources supersede most of the previous large surveys in HI galaxies, HI imaging, and pulsar search.

In summary, deploying the 19-beam in a drift scan is a currently feasible and near optimum survey strategy, utilizing the amazing gain of FAST and minimizing complexities in system control and RFI mitigation for best system performance.

## IV. A Commensal Radio Astronomy FAST Survey and ITS Technical Challenges

We hereby designate our drift survey plan as a Commensal Radio Astronomy FasT survey (CRAFTS). The motivation for CRAFTS is straightforward, considering the unique advantages and limitations of FAST as discussed in the previous session. The push for commensality was also one of the key recommendations from the 2[nd] Frontiers in Radio Astronomy Symposium [7]. In the initial discussions of the Arecibo L-band Focal Plan (ALFA) survey plans, commensality was also one of the main focus. In practice, though, "piggy-back" mode was accomplished instead. The main difference lies in the principles of decision. In piggy-back mode, the main survey, a.k.a the 'pig', dictates the survey design, which requires the 'backer' to adopt. In a commensal mode, multiple science goals co-exist and force compromises, wherever necessary. A primary example of a successful 'backer' is the Galactic ALFA project (GALFA) [8]-[10]. GALFA scientists managed to combine data from their specifically designed basket-weaving scanning mode and those taken in piggy-back mode along with various other projects, such as the Arecibo Legacy Fast ALFA Survey (ALFALFA, [11]). To piggy-back HI imaging onto a pulsar-search survey such as the P-ALFA [12], however, has proven to be exceedingly difficult. All major radio observatories, such as Arecibo, Effelsberg, GBT, and Parkes, have conducted multiple large surveys to observe pulsar and HI. However, a large scale commensal survey, particularly between HI imaging and pulsar search, remains elusive.

The drift scan mode of CRAFTS is necessitated by the unique design of FAST and facilitated by its world-leading sensitivity. The majority of past pulsar surveys were carried out by staring and mosaicking. The most successful pulsar survey, namely, the Parkes Multi-beam Pulsar Survey (PMPS, [13]-[14]) covered the southern Galactic plane by tiling the plane area in a beam-spacing pattern with the Parkes 13-beam array. The staring time is about 50 minutes to achieve a coherent relatively long time-series to allow for searching for periodical signal in the power spectrum. Individual beam has no overlap with each other. It is not required and not feasible to combine data from different beams, nor from different pointing, because their phases cannot be connected. Such data cannot be used to construct a high-quality HI image, even with an additional spectral backend operating simultaneously, due to its under-sampling in the spatial domain and the gain variation in both the time and spatial domain.

We discuss here an innovative and experimental calibration technique, specifically designed for CRAFTS. A common technique to characterize the short-time-scale gain variation is to inject an electronic noise signal, usually referred to as a CAL, every second or so. The strength of the CAL is determined by the requirement of having enough S/N to measure the system temperature. To optimize such measurement, usually the CAL is on half of the integration time. Although the CAL is relatively weak, say 1 K, compared to the system noise level (~20 K), it is much stronger than most pulses from neutron stars. For example, the PMPS flux limit was roughly 0.1 mJy, which corresponds to only 1.6 mK even with the much larger gain of FAST. Such a CAL will also populate the power spectrum with its harmonics, rendering pulsar search difficult, if not impossible (Figure 4).

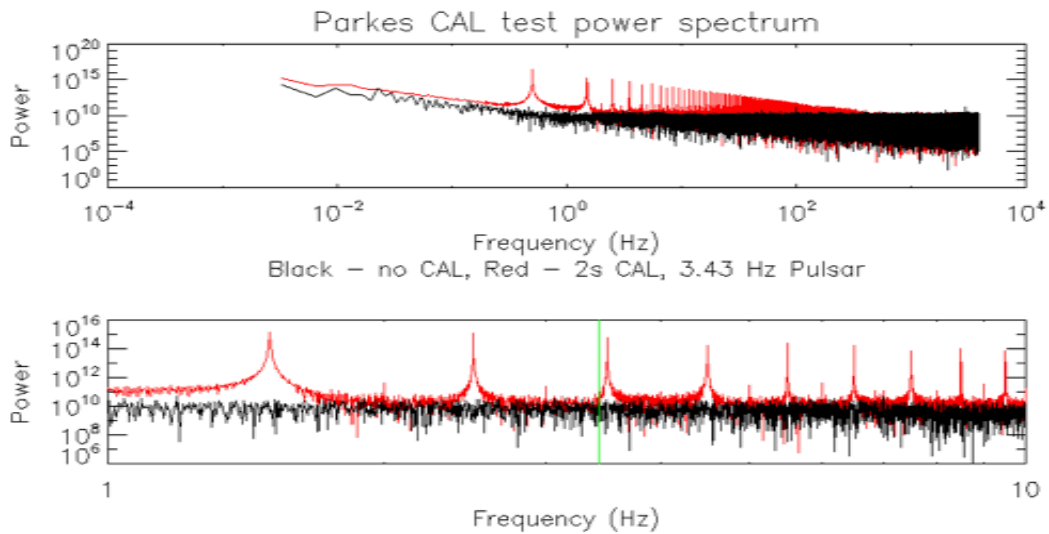

**Fig. 4** *Results from a CAL injection experiment at Parkes with pulsar search mode. The experiment was carried by M. Krco with great help from G. Hobbs and other CSIRO staffs. CAL signal clearly dominated the resulting power spectrum.*

Our current proposal is to inject the CAL at the data sampling rate. The plan is also to save a copy of the CAL signal temporarily. While the CAL data stream is being integrated to flag a Tsys at about 1 s interval, it will also be subtracted from the pulsar data stream at the sampling rate. The result pulsar stream will be much cleaner. Even if the subtraction leaves residues, such a fast injection rate constrains the CAL power and its harmonics into much below 1 ms, which is shorter than any of the known or expected pulsar period. Such a calibration scheme works in simulation and is being tested in the commissioning process. Once successful, it will be key innovation to facilitate CRAFTS.

CRAFTS uses at least four backends for pulsars, HI galaxies, HI imaging, and fast radio bursts (FRB). A possible fifth search-for-extraterrestrial-intelligence (SETI) spectrometer is being discussed. The sampling rate of pulsars is 100 milliseconds. For HI observations, the sampling rate is 1 second. The FRB backend has a real-time triggering mechanism, which saves baseband data in a one-minute ring buffer when triggered.

The international Galactic plane surveys (VGPS [15], CGPS [16], SGPS [17]) have velocity resolution of about 1 km/s. The GALFA survey's resolution is much higher at 0.18 km/s. The fine velocity resolution is driven by absorption features rather than Galactic emission. Self-absorption features [18] can be as narrow as 1 km/s. The so-called HI Narrow Self-absorption (HINSA, [19]) features can even approach the thermal linewidth of a dark cloud, which is only about 0.3 km/s. To preserve CRAFTS' capability at revealing cold HI throughout the Galaxy, a 1 KHz frequency resolution is required. For a 20 MHz band that is sufficient to both cover nearby Galaxies and resolve MilkyWay cold gas, a total of 20 K channels are needed. The modern SETI backend, only a possible addition at this stage, could have orders of magnitude more channels. The number of channels for each backend thus ranges between 4 K to 20 K, with HI imaging having the highest number of channels.

Given the vast difference in sampling rate, the data rate and volume will be dominated by pulsars. For 8 bit sampling of the FAST backends, 100 μs time sampling, 4 K channels, 2 polarizations, 19 beams, the data rate will amount to 1.6 GB/s, 5.8 TB/h, and 144 TB/day. The annual data volume will depend upon the operational conditions and the time allocation between surveys and PI-led programs, but falls within the 10 to 20 PB range. The Large Synoptic Survey Telescope (LSST) is the poster child for peta-scale and time-domain astronomy, with an expected data rate of 15 TB/day and 60 PB total for a 10 year span according to https://www.lsst.org. It becomes impossible at this scale to require comprehensive human check of the majority of the data. The planned LSST processing power is 150 Tflops. CRAFTS pulsar search will generate tens to hundreds of thousands of pulsar candidates from each 24-hour scan. These candidates need to be sorted quickly to allow planning for confirmation observation and follow up timing. The processing time versus data taking time has to be similar to avoid cumulative lag. Our testing with the PRESTO pipeline [20] shows that a near-real-time processing of CRAFTS pulsar data stream requires, at the minimum, 100 Tflops. In short, CRAFTS will have to supersede LSST in terms of instantaneous data rate and rival LSST in terms of total volume and processing power. CRAFTS will present another landmark in peta-scale astronomy, with challenges necessarily associated with big data.

The key challenge and the primary consideration in seeking big-data solutions is money. The ¥1.15B project budget for FAST includes exactly ¥0 for scientific data processing. There is not nearly enough physical space on site for necessary data processing equipments. The usual expectation of an annual operational budget at ~10% construction budget level does not scale properly for peta-scale astronomy. For example, about 1 full time employee (FTE) will be needed to maintain each PB of data, in terms of archiving, accessing, and processing. The FAST operation budget is not expected to support more than a few FTE for purely data related tasks. We are at least one order of magnitude short if the CRAFTS data is to be dealt with by a single entity.

Our current data plan is to purchase a 100 Gb/s fiber link to transit filterbank data in real time from the site to two early science centers near Guiyang city, Guizhou. One center is being operated by the Guizhou Normal University Computer Center (led by Prof. Xie, Xiaoyao and Prof. Liu, Zhijie), with 2

PB and 20 Tflops capabilities. The other will be a rental contract with the local telecom company, with 10 PB and 100 Tflops. These two centers are to be equipped with our own database-based pulsar search pipeline with AI ranking [21]. To handle CRAFTS data beyond the first full year of operation, efforts are spent on multiple fronts. First, we will build our own data center within the newly established Guizhou Observatory, Chinese Academy of Sciences, aiming for 10 PB and 100 Tflops. Second, collaborative data centers will be supported jointly by and located in Nanjing University, Yunnan University, Beijing Normal University, Peking University, and etc., each at PB scale. Third, commercial cloud computing, such as the Ali-cloud, is being tested with expected ensuing negotiations. In summary, the data processing of CRAFTS will necessarily be accomplished by a distributive, collaborative, and cloud-like infrastructure.

## V. THE MULTI-BEAM SYSTEM AND THE SURVEY DESIGN

The CRAFTS observations will be made using the FAST L-band Array of 19 feed-horns (FLAN, [22]). This cryogenic multi-beam receiver is similar in design to the Parkes 13 beam [23] and the ALFA 7 beam [24] multi-beam receivers. Apart from the increased number of beams, the main difference between FLAN and these receivers is that the bandwidth has been increased to cover the 1.05-1.45GHz frequency range. Typical multi-beam receivers designed for prime focus reflectors have been constrained to bandwidth ratios of less than 1.2:1. This is largely due to the limitations placed on the feed horns by their small spacing. The size limitation imposed on the feed horns by their spacing results in over-illumination of the reflector at low frequencies. The unique structure of FAST is advantageous in this regard because illumination outside the parabolic portion of the reflector falls not on the ground but on the spherical section of the reflector. As the spherical section of the reflector reflects the low temperature radiation of the sky it contributes only a fraction of the spill-over noise that would be seen in telescopes like the Parkes Radio Telescope. This tolerance to a wide illumination pattern allows the bandwidth to be increased to 1.4:1 and high aperture efficiency is possible across the whole band. The polarizations of each beam are separated using a quad-ridged ortho-mode transducer (OMT). Mechanical size and weight constraints limit the length of this OMT to approximately 1.5 wavelengths at 1.05GHz. Despite this limitation a return loss of greater than 20dB has been achieved across the whole band. The gain is degraded by less than 1dB for the beam at the edge.

Careful attention has been paid to maintaining symmetry throughout the OMT. Symmetry is broken only in the probe area of the OMT and in this area dimensions are reduced as much as possible to avoid the coupling associated with this asymmetry. By reducing these asymmetries excitation of the trapped $TE_{11}$ mode has been reduced to a level such that the polarisation cross coupling only rises to -40dB at worst. The cryogenic portion of the receiver is cooled by three CTI 1050 cryocoolers. The OMTs are cooled to 60K, the low noise amplifiers to 15K and the remainder of the feed system is at ambient temperature. The measured noise temperature of the receiver is 8K±2K.

The layout of FAST L-band Array of 19 feed-horns (FLAN) is similar to that of the Parkes 13-Beam, with two concentric rings of feeds surrounding the center one (Figure 5). The beam pattern on the sky also follows those of Parkes 13-beam and ALFA in that there is one beam spacing between beams. 3dB overlapping can be expected after tessellation mode. As discussed earlier, PMPS used mosaicing to cover the Galactic plan for pulsar search, while HI observations prefer drift-scan for better spatial sampling.

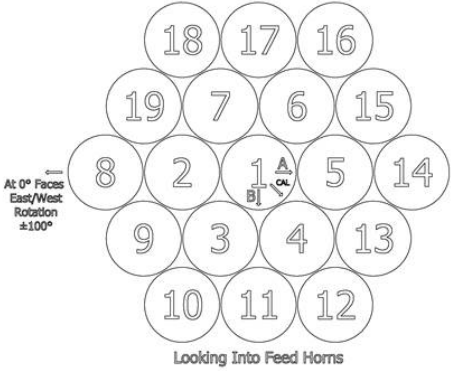 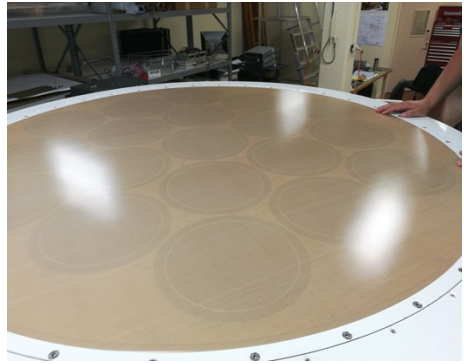

**Fig. 5** *FAST L-band Array of 19 feed-horns (FLAN). Left: Feed layout and numbering Right: A Picture taken in May of 2017 in Marsfield, Sydney to illustrate the FLAN's physical scale (with its cover and encasing).*

The FAST focal cabin is driven by 6 cables along a curved focal plane. Such a mechanism allows for a fixed rotation angle (counted clockwise from north) of the beam pattern at a certain declination (DEC) on the sky while the drifting drags the beam through right ascension (RA). The spacing between beam patterns depends upon the rotation angle (Figure 6). By rotating the array by 23.4 degrees, the spacing between the inner 17 beams will be super-Nyquist sampling, namely, smaller than half the FWHM. The spacing between two outer beams and their respective adjacent beams is twice as large. Those spacing will be filling up by moving the whole array in DEC by 21.9' for the next scan. In this pattern, CRAFTS need 220 days to cover the FAST sky between-14 deg to +66 deg of DEC.

## VI.   CRAFTS' SCIENCE POTENTIAL

FAST covers continuously the EM spectrum from 70 MHz to 3 GHz, which include several key science targets, namely, radio pulsars, the 21 cm HI hyperfine structure line (both in galaxies and in the Milky Way, radio continuum, recombination lines, and molecular spectral lines including masers. Correspondingly, the overall science goals of include neutron star physics, low frequency gravitational wave detection and gravity tests, cosmology and galaxy evolution, Galactic structure, interstellar medium and star formation, VLBI, and SETI [1], [2], [7]. Quantified science expectations have been published by the international radio astronomy communities in various medium, e.g. pulsar surveys [25]-[27], HI galaxy detection [28], HI intensity mapping [29], HI galaxy absorption [30], OH mega-masers [31]-[32], gamma-ray bursts (GRB, [33]), fast radio bursts (FRB, [34]), and etc.

CRAFTS is designed to be a quick-look and 'discovery' survey, rather than achieving best the possible results with FAST for any particular objective in other specially designed modes, such as long time tracking and basket-weaving scanning. For example, CRAFTS will cover M31, but will not be sensitive enough to detect normal pulsars, which would be a major discovery given that no radio pulsar has been seen in any spiral galaxy other than our own. CRAFTS, however, is capable of detecting giant pulses, which will also be a significant discovery. The outlook of such programs is highly speculative. CRAFTS is thus designed to preserve as much discovery space as possible, limited by the data volume and data rate issues discussed earlier.

## VII.   SUMMARY

If successful, CRAFTS expects to discover more than a thousand pulsars, close to 200 of which will be MSPs, detect hundreds of thousands HI galaxies, obtain 10 billion voxels in HI images, and discover a few 10s of FRBs. The data rate, calibration, scan design, etc. of CRAFTS pose significant challenges. The commensality has long been desired, but remains unprecedented. Combined with the also unprecedented sensitivity of FAST, such a commensal survey should make fundamental contribution to studies of pulsars, galaxies, ISM, and transients and foster potential breakthroughs.

## ACKNOWLEDGMENT

This work is supported by the National Key R&D Program of China No. 2017YFA0402600 and the International Partnership Program No. 114A11KYSB20160008 and the Strategic Priority Research Program No. XDB23000000 of the Chinese Academy of Sciences.

## REFERENCES

[1] R. Nan, D. Li, C. Jin, Q. Wang, L. Zhu, W. Zhu, et al., "The Five-hundred-meter Aperture Spherical radio Telescope (FAST) project," International Journal of Modern Physics D, vol. 20, no. 6, pp. 989-1024, Feb. 2011.

[2] D. Li and Z. Pan, "The Five-hundred-meter Aperture Spherical Radio Telescope project," Radio Science, EGU Publication, vol. 51, no. 1060, pp. 1-5, May. 2016.

[3] P. Jiang, R. Nan, L. Qian, Y. Yue, "Studying solutions for the fatigue of the FAST cable-net structure caused by the process of changing shape," Research in Astronomy and Astrophysics, vol. 15, no. 10, pp. 1758-1772, Jan. 2015.

[4] P. Jiang, Q. Wang and Q. Zhao, "Optimization and analysis on cable net structure supporting the reflector of the large radio telescope FAST," Applied Mechanics and Materials, vol. 94, pp. 979-982, 2011.

[5] R. Yao, W. Zhu and P. H., "Accuracy analysis of stewart platform based on interval analysis method," Chinese Journal of Mechanical Engineering, vol. 26, no. 1, pp. 29, 2013.

[6] J. Sun, R. Nan, W. Zhu, H.J. Kärcher and H. Li, Modeling, Systems Engineering, and Project Management for Astronomy III, 70171L, 2008.

[7] L. Qian and D. Li, Conference Proc., Astronomical Socirty of the Pacific Conference Series, vol. 502 (http://aspbooks.org/a/volumes//502), 2016.

[8] P.F. Goldsmith and Bulletin of the American Astronomical Society, 36, 2005.

[9] J.E.G. Peek, C. Heils and K.A. Douhlas et al., AAS Meeting, 194, 2011.

[10] J.E.G. Peek and M.E. Putman, AAS Meeting, 214, 2009.

[11] M. Haynes et al., "ALFALFA HI data stacking i. does the bulge quench ongoing star formation in early-type galaxies?," MNRAS, vol. 411, pp. 993-1013, 2011.

[12] M. Kaspi et al. AAS Meeting, 2012.

[13] R.N. Manchester, A.G. Lyne and N. D'Amico, et al., "The parkes Southern pulsar Survey — I. Observing and data analysis systems and initial results," MNRAS, vol. 279, pp. 1235-1250, 1996.

[14] D.R. Lorimer, A.J. Faulkner, A.G. Lyne, R. Manchester, M. Kramer and M.A. McLaughlin, et al., "The parkes multibeam pulsar survey – vi. discovery and timing of 142 pulsars and a galactic population analysis," MNRAS, vol. 372, pp. 777-800, 2006.

[15] J.M. Stil, A.R. Taylor and J.M. Dickey, "The VLA galactic plane survey," ApJ, vol. 132, pp. 1158 -1176, 2006.

[16] K.A. Douglas and A.R. Taylor, "The view of the interstellar medium with the canadian galactic plane survey," part of the Astrophysics and Space Science Library book series (ASSL), vol. 315, pp. 43-46, 2004.

[17] N.M. McClure-Griffiths, J.M. Dickey and B.M. Gaensler, et al., "The southern galactic plane survey: h i observations and analysis," ApJS, vol. 158, no. 2, pp. 178-187, 2005.

[18] S.J. Gibson, A.R.Taylor and P.E. Dewdney, ASP Conference Series, vol. 168, 1999.

[19] D. Li and P.F. Goldsmith, "HI narrow line absorption in dark clouds," ApJ, vol. 585, pp. 823-839, Jun. 2003.

[20] S.M. Ransom, "New search techniques for binary pulsars," Ph.D thesis, 2001.

[21] W.W. Zhu, A. Berndsen and E.C. Madsen et al., "Searching for pulsars using image pattern recognition," ApJ, vol. 781, no. 117, pp. 1-12, Feb. 2014.

[22] A. Dunning, M. Bowen, S. Castillo et al., in General Assembly and Scientific Symposium of the International Union of Radio Science (URSI GASS—XXXII), 2017.

[23] L. Staveley-Smith, W.E. Wilson, T.S Bird et al. 1996, Publ. Astron. Soc. Aust., 13, 243-248

[24] G. Carrad, P. Sykes and G. Moorey, in Workshop on Applications of Radio Science (WARS2006), 2006.

[25] R. Smits, D.R. Lorimer and M. Kramer et al., "Pulsar science with the five hundred metre aperture spherical telescope," A&A, vol. 505, pp. 919-926, Feb. 2009.

[26] Y. Yue, D. Li and R. Nan, "FAST low frequency pulsar survey," IAU Symposium, no. 291, pp. 57-59, Mar. 2013.

[27] G. Hobbs, S. Dai, R.N. Manchester et al., "The Role of FAST in Pulsar Timing Arrays," to be published in a special issue on FAST in RAA, arXiv: 1407.0435.

[28] A. Duffy, R. Battye, R.D. Davies, A. Moss and P. Wilkinson, SISSA, Proceedings of Science, vol. 91, 2007

[29] M.A. Bigot-Sazy, Y.Z. Ma, R.A. Battye, et al., Astronomical Socirty of the Pacific Conference Series, vol. 502, pp. 41, 2016.

[30] H.R. Yu, T.J. Zhang and U.L. Pen, "Method for direct measurement of cosmic acceleration by 21-cm absorption systems," Phys. Rev. Lett., vol. 113, no. 041303, pp. 1-4, July. 2014.

[31] J.S. Zhang, J.Z. Wang, D. Li and Q.F. Zhu, IAU Symposium, vol. 35, pp. E85, 2006.

[32] J.S. Zhang, D. Li and J.Z. Wang, IAU Symposium, vol. 287, pp. 350, 2012.

[33] Z.B. Zhang, S.W. Kong, Y.F. Huang, D. Li, L.B. Li, "Detecting radio afterglows of gamma-ray bursts with FAST," Research in Astronomy and Astrophysics, vol. 15, no. 2, pp. 237-251, 2015.

[34] L.B. Li, Y.F. Huang, Z.B. Zhang and D. Li, et al., "Intensity distribution function and statistical properties of fast radio bursts," Research in Astronomy and Astrophysics, vol. 17, no. 1, pp. 6-12, 2017.